\documentclass[epj]{svjour}
\usepackage[english]{babel}
\usepackage{graphicx}
\usepackage{dcolumn}
\usepackage{bm}
\usepackage{color}
\usepackage{subfigure}
\usepackage[ansinew]{inputenc}
\usepackage{amsfonts}
\usepackage{amsmath}
\usepackage{amssymb}
\usepackage{slashed}

\def\e{\mathrm{e}}

\def\e{\mathrm{e}}
\def\i{\mathrm{i}}
\def\d{\mathrm{d}}

\begin{document}
\title{Huygens' principle and Dirac-Weyl equation}
\author{Saverio Pascazio\inst{1,2,3} \and Francesco V. Pepe\inst{2,4} \and Juan Manuel P\'{e}rez-Pardo \inst{5}
}                     
%
%
\institute{Dipartimento di Fisica and MECENAS, Universit\`{a} di Bari, I-70126 Bari, Italy \and INFN, Sezione di Bari, I-70126 Bari, Italy
\and Istituto Nazionale di Ottica (INO-CNR), I-50125 Firenze, Italy \and Museo Storico della Fisica e Centro Studi e Ricerche ``Enrico Fermi'', I-00184 Roma, Italy
\and Universidad Carlos III de Madrid, 28911 Madrid, Spain}
\date{Received: date / Revised version: date}
%
\abstract{
We investigate the validity of Huygens' principle for forward propagation in the massless Dirac-Weyl equation. The principle holds for odd space dimension $n$, while it is invalid for even $n$. We explicitly solve the cases $n=1,2$ and $3$ and discuss generic $n$. We compare with the massless Klein-Gordon equation and comment on possible generalizations and applications.%
} 
\maketitle
\section{Introduction}
\label{intro}
Every point on the wave front of a propagating wave is a source of secondary wavelets, which spread forward at the same speed as the source wave. The wave front at later times is then given by the surface tangent to the secondary wavelets. This principle was proposed by Christiaan Huygens in 1678, to explain the laws of reflection and refraction. It was used again more than a century later, in 1816, by Augustin-Jean Fresnel, to interpret the diffraction effects that occur when visible light encounters slits, edges and screens. 

The principle provides crucial insight into the nature of wave propagation and it is a milestone in the physics of undulatory phenomena \cite{BornWolf}. For this reason, its universal validity is usually taken for granted. However, yet one century later, Jacques Hadamard noticed that Huygens' principle is valid only when waves propagate in an odd number $n$ of spatial dimensions, with the notable exception of $n=1$ \cite{Hadamard}.

The mathematical formulation of the principle is embodied in the explicit formulas that give the solution of the Cauchy initial value problem for the wave equation. The solution at time $t$ and position $\bm x$ is expressed as an integral that involve the wave and its derivative at time $t=0$. For odd $n \geq 3$, the integration domain is a (traveling) \emph{surface}. For even $n$, the integration domain is a \emph{volume}, enclosed by a traveling surface. Take for example as initial condition a Dirac delta function at the origin in homogenous space.
For odd $n \geq 3$, the solution at time $t$ is non-vanishing on a \emph{sphere} of radius $r=ct$, $c$ being the speed of propagation; for even $n$ the solution at time $t$ is non-vanishing in a \emph{ball} of radius $r=ct$. This is the reason why we hear a sound only once, but we observe a ``decaying trailing edge" when we throw a stone in a pond. The beautiful mathematical details of these phenomena were discussed by Hadamard by making use of his ``method of descent" for solving partial differential equations. 
It has been used since to solve a variety of differential equations and Cauchy initial value problems \cite{zac,evans}. 

Both quantum mechanics and quantum field theory make use of wave equations in their formulation. It is therefore interesting to ask whether Huygens' principle holds for the seminal equations that are the backbone of these theories. The Schr\"odinger equation, being non-relativistic {and intrinsically dispersive}, does not admit a satisfactory formulation of this question. What about the Dirac equation? 

The aim of this article is to discuss the validity of the Huygens' principle for the Dirac equation. This problem was investigated in the mathematical literature, where general results were given, for different dimensionality \cite{Gunther,chalub,chalub2}. {Our objective is twofold: we will discuss the problem in terms of the propagator (in the position and momentum representations) and will endeavor to analyze the physical features of wave propagation in terms of its singularities, highlighting the differences between even and odd space dimensions. This article contains also a tutorial part, whose objective is to bridge the gap between the approach we adopt, based on the Green function, and the use of initial conditions (Cauchy problem for homogeneous differential equations). For the sake of clarity, we shall always give the explicit solutions.

We shall focus on wave propagation \textit{forward in time}, so that the Dirac equation will be considered at the classical level. Information on forward propagation is encoded in the \emph{retarded} Green functions, and is fully characterized by the treatment of its singularities. Unlike Feynman's choice, that emerges in quantized field theories and that we plan to analyze in a future article, the retarded choice propagates both the positive- and negative-energy parts of the initial condition forward in time.} We will restrict our analysis to the massless case and will find that the Dirac-Weyl equation inherits the main features of the Klein-Gordon wave equation with the notable difference that Huygens' principle is valid \emph{also} for $n=1$. We will give general arguments, valid for all $n$, and explicit expressions for $n\leq 3$.

\section{Massless Klein-Gordon and Dirac-Weyl equations}
\label{massless}
We will work in $n$ space dimensions, with 
 $x=(x^\mu)=(t,\bm x)$, the index $\mu$ running from 0 (time) to $n$ (space), and $(p^\mu)=(p^0,\bm p)$. 
 Einstein summation convention over repeated indices is implied, and $\hbar = c =1$.
Let $G_\textrm{R}$ be the propagator of the massless Klein-Gordon (KG) equation 
\begin{equation}
\label{KGmassless}
\partial^\mu \partial_\mu G_\textrm{R} (x) = \Box G_\textrm{R} (x) = - \i \delta (x) ,
\end{equation}
where the subscript $_\textrm{R}$ stands for ``retarded", propagating signals forward in time. Here and in the following, $(\partial_{\mu})=(\partial_t,\bm{\nabla})$, and indices are raised and lowered by the metric tensor $\left(\eta_{\mu\nu}\right)=\left(\eta^{\mu\nu}\right)=\mathrm{diag}(+1,-1,\dots,-1)$. The Fourier transform of \eqref{KGmassless} reads 
\begin{equation}
\label{propKGR}
G_\textrm{R} (x) = \int \frac{\d^{n+1} p}{(2 \pi)^{n+1}} \, \e^{-\i p_{\mu}  x^{\mu}} \frac{\i}{p_{\mu}p^{\mu} + \i \epsilon \, \mathrm{sgn}(p_0)}
\end{equation} 
where $ \mathrm{sgn}(k)=1$ for $k>0$ and $-1$ for $k<0$, and one finds
\begin{eqnarray}
\label{propKGexpl}
G_\textrm{R} (x)  & = & -\i \int \frac{\d^n {\bm p}}{(2 \pi)^n} \, 
\frac{\sin (p t)}{p} 
 \e^{\i {\bm p} \cdot {\bm x}} \, \theta (t),
\end{eqnarray}
with $\theta$ the Heaviside step function and $p:=|\bm{p}|$.

Consider now the massless Dirac equation in $n+1$ dimensions
\begin{equation}
\label{Dmassless}
\i \slashed{\partial} \, \psi = \i \gamma^\mu \partial_\mu \psi = 0 ,
\end{equation}
where the gamma matrices obey Clifford's algebra 
\begin{equation}
\label{clifford}
\{\gamma^\mu, \gamma^\nu  \}= 2\eta^{\mu\nu}.
\end{equation}
We shall only consider the massless case, in which the Dirac equation splits into uncoupled (Weyl) equations. The {retarded Dirac} propagator, satisfying
\begin{equation}\label{Dmasslessprop}
\i \slashed{\partial} D_{\textrm{R}}(x) = \i \gamma^0 \delta (x)
\end{equation}
generally reads \cite{Peskin}
\begin{eqnarray}
\label{dpropform}
D_{\textrm{R}} (x) = \i \slashed{\partial} \gamma^0 G_\textrm{R} (x) ,
\end{eqnarray} 
where $\gamma^0$ has been introduced for convenience. This formally shows that, if the boundary conditions are properly handled, the Dirac propagator inherits the main features of its KG counterpart. Since this is not the Feynman prescription for the propagator \cite{Peskin,Feynman}, let us give the explicit expression:
\begin{eqnarray}
D_{\textrm{R}} (x)  & = & \int \frac{\d^{n+1} p}{(2 \pi)^{n+1}} \e^{- \i p_{\mu} x^{\mu} } \frac{\i \slashed{p} \gamma^0}{ p_{\mu}p^{\mu} + \i \epsilon \,\mathrm{sgn}(p_0)} 
= \int  \frac{\d^n \bm p}{(2 \pi)^n} \, 
\e^{\i {\bm p} \cdot {\bm x}} \e^{-\i(\gamma^0\bm{\gamma}\cdot\bm{p} -\i\epsilon)t} \, \theta(t) ,
 \label{propDexpl}
\end{eqnarray}
where
\begin{equation}
\e^{-\i\gamma^0\bm{\gamma}\cdot\bm{p} t} = \cos (pt) -\i 
\frac{\gamma^0  \bm{\gamma} \cdot \bm{p}\, t}{p} \sin (pt) .
\end{equation}
This expression is valid for all $n$.

\section{Forward propagators and initial value problem}
\label{invalue}

The retarded Green functions that satisfy equations like \eqref{KGmassless} and \eqref{Dmasslessprop} can generally be used to find the particular solutions of inhomogeneous equations. In the Dirac case, the equation
\begin{equation}
\i \slashed{\partial} \psi(x) = j(x),
\end{equation}
with $j$ a given external source, is solved by
\begin{equation}\label{inhom}
\psi(x) = \psi_{\textrm{h}}(x) - \i \int \d^{n+1}y D_{\textrm{R}}(x-y) \gamma^0 j(y) ,
\end{equation}
with $\i\slashed{\partial}\psi_{\textrm{h}}(x)=0$. However, since the Dirac equation is first-order, the forward propagator is useful also in the solution of initial value problems. If one is interested in the solution of the \emph{free} Cauchy problem
\begin{equation}\label{Cauchy}
\left \{ \begin{array}{l} \i \slashed{\partial} \psi(x) = 0 \\ \\ \psi(0,\bm{x})=\psi_0(\bm{x}) \end{array} \right. 
\end{equation}
at $t>0$, then one can associate to $\psi$ the auxiliary function $\tilde{\psi}(x)=\psi(x)\theta(t)$, coinciding with the solution at the times of interest, and satisfying the \emph{inhomogeneous} equation
\begin{equation}
\i \slashed{\partial} \tilde{\psi}(x) = \i \gamma^0 \psi_0(\bm{x}) \delta(t) ,
\end{equation}
whose general solution has the form \eqref{inhom}, with $j(x)=\i\gamma^0\psi_0(\bm{x}) \delta(t)$. Moreover, since from Eq.\ \eqref{propDexpl}
\begin{equation}
\lim_{t\to 0^+} D_{\textrm{R}} (t, \bm{x}) = \delta(\bm{x}),
\end{equation}
no solution $\psi_{\textrm{h}}$ of the homogeneous equation is needed to match the initial condition, and
\begin{equation}
\label{forward}
\tilde{\psi}(t,\bm{x}) = \psi(t,\bm{x}) \theta(t) = \int \d^n\bm{y} D_{\textrm{R}}(t,\bm{x}-\bm{y}) \psi_0(\bm{y}) 
\end{equation}
coincides with the (unique) solution of the Cauchy problem \eqref{Cauchy} for $t>0$. 

Before explicitly looking at the cases $n=1,2$ and $3$, let us comment on the relation with the Feynman propagator. The retarded Green's function $D_{\textrm{R}}$ in Eq.~\eqref{forward} propagates any initial spinor $\psi_0(\bm{x})$ forward in time. On the other hand, only positive-frequency components are propagated forward \emph{\`a la} Feynman: thus, if the initial spinor is made of positive-frequency components only, Feynman and retarded propagation are indistinguishable for $t>0$.

\subsection{1+1 dimensions}
\label{sec:11dim}
We shall look at this case in some detail, as it serves as an introduction to the techniques illustrated in Secs.\ 
\ref{massless}-\ref{invalue} \emph{and} it gives results that are at variance with Klein-Gordon.
Let $\gamma^0 = \sigma^3$ and $\gamma^1 = - \i \sigma^2$, where the $\sigma$'s are the Pauli matrices.
The Dirac equation (\ref{Dmassless}) is written in terms of a two-component spinor $\psi$ and its propagator is a $2\times 2$ matrix $D_{\textrm{R}}^{(1)}$, where 
the superscript labels the space dimension. 

Since $(x^{\mu}) = (t,x)$ for $n=1$, one easily gets  
\begin{align}
D_{\mathrm{R}}^{(1)} (t,x) & = \left( \begin{matrix} \partial_0 & \partial_1 \\ \partial_1 & \partial_0 \end{matrix}  \right) \i G_\textrm{R}^{(1)}(t,x) = \frac{1}{2} \left( \begin{matrix} \delta (x+t) + \delta (x-t) & \delta (x+t) - \delta (x-t) \\ \delta (x+t) - \delta (x-t) & \delta (x+t) + \delta (x-t) \end{matrix} \right)
 \label{D1}
\end{align} 
by direct computation from \eqref{propDexpl}. This clearly shows that Huygens' principle is valid, a result already known in the matematical literature \cite{chalub}. It is interesting to look at the expression of the time-evolved Dirac field. One obtains from Eq.\ (\ref{forward})
\begin{eqnarray}
\psi (t,x) &=& \int \d y D_{\textrm{R}}^{(1)} (t,x-y) {\phi_0 (y) \choose \chi_0 (y)}
= \frac{1}{2} 
{(\phi_0 + \chi_0 )(x+t) + (\phi_0 - \chi_0 )(x-t) \choose (\phi_0 + \chi_0 )(x+t) - (\phi_0 - \chi_0 )(x-t)} , 
\label{psi1} 
\end{eqnarray} 
where $\phi_0$ and $\chi_0$ are the initial conditions of the two-component spinor.

This result is seemingly in contrast with the solution of the massless KG equation, reading
\begin{equation}
\label{eq:KGsolution}
\phi(t,x) = \frac{1}{2}(f(x+t) + f(x-t)) + \frac{1}{2}\int^{x+t}_{x-t} g(y)\d y,
\end{equation}
with initial conditions $\phi(0,x) = f(x)$ and $\partial_t\phi(t,x)|_{t=0} = g(x)$, where Huygens' principle is not valid due to the 1D volume integral [the last addendum in (\ref{eq:KGsolution})]. The relation between the structure of the solutions (\ref{psi1})-(\ref{eq:KGsolution}) and Huygens' principle, valid only for the Dirac equation, can be unveiled in two alternative ways. First, consider two solutions $\phi_1$ and $\phi_2$ of KG, expressed in the form \eqref{eq:KGsolution} through pairs of functions $(f_1,g_1)$ and $(f_2,g_2)$, respectively. The two-component object $\Phi=(\phi_1\,\,\phi_2)^{\mathrm{T}}$ is generally not a solution of massless Dirac equation. However, the spinor 
\begin{equation}
\psi(t,x)=\slashed{\partial}\Phi(t,x) = \left( \begin{matrix} \partial_t \phi_1(t,x)- \partial_x \phi_2(t,x) \\ \partial_x \phi_1(t,x) - \partial_t \phi_2(t,x)                
\end{matrix} \right)
\end{equation}
satisfies at the same time the Dirac equation and Huygens' principle, since it depends only on the values of the one-variable functions $f'_j(x\pm t)$ and $g_j(x\pm t)$, with $j=1,2$, on the light cone. 

An alternative way to understand the emergence of light-cone propagation in passing from KG to Dirac is the equivalence between the Dirac equation for the spinor $\psi=(\psi_1\,\,\psi_2)^{\mathrm{T}}$ and two KG equations for its components, with initial conditions $\psi_j(0,x)=\psi_j^0(x)$ \emph{and}
\begin{equation}
\partial_t \psi_1(t,x)|_{t=0} = \partial_x \psi_2^0(x), \qquad \partial_t \psi_2(t,x)|_{t=0} = \partial_x \psi_1^0(x) .
\end{equation}
Thus, out of the four independent functions needed to define the Cauchy problems for the two spinor components obeying the KG equation, there are only two that can be chosen independently, in order to satisfy the Dirac equation. Moreover, the initial time derivatives coincide with spatial derivatives, which simplifies the bulk integral in \eqref{eq:KGsolution} into a difference of functions on the light cone.

\subsection{2+1 dimensions}
\label{sec:21dim}
In two spatial dimensions, the Dirac matrices can be chosen as $\gamma^0=\sigma^3$, $\gamma^1=-\i\sigma^2$ and $\gamma^2=\i\sigma^1$
and the Dirac spinor is again 2-dimensional. The forward propagator in momentum space reads
\begin{equation}
\e^{-\i\gamma^0 \bm{\gamma}\cdot\bm{p} t} = \left( \begin{matrix} \cos(pt) & (\i p^1 + p^2) \frac{\sin(pt)}{p} \\ (\i p^1 - p^2) \frac{\sin(pt)}{p} & \cos(pt) \end{matrix}  \right) ,
\end{equation} 
while its Fourier transform can be expressed as
\begin{equation}
D_{\textrm{R}}^{(2)}(t,\bm{x}) = \left( \begin{matrix} \partial_0 & \partial_1 - \i \partial_2  \\ \partial_1 + \i \partial_2 & \partial_0 \end{matrix}  \right) \i G_\textrm{R}^{(2)}(t,\bm{x})
\end{equation}
where 
\begin{eqnarray}
G_\textrm{R}^{(2)} (t,\bm{x}) & = & -\i \int \frac{\d^2p}{(2\pi)^2} \frac{\sin(pt)}{p} \e^{\i\bm{p}\cdot\bm{x}}  =  -\i \int_0^{\infty} \frac{\d p}{2\pi} J_0(pr) \sin(pt),
\end{eqnarray}
with $r=|\bm{x}|=\sqrt{(x^1)^2+(x^2)^2}$ and $J_0$ a Bessel function of the first kind. The integral can be evaluated by adding small imaginary parts $\pm i\epsilon$ to time $t$: in the limit $\epsilon\to 0$, it reads
\begin{equation}
G_\textrm{R}^{(2)} (t,\bm{x}) = \frac{-\i}{2\pi \sqrt{t^2-r^2}} \theta (t-r).
\label{GR2}
\end{equation}
Notice that contributions outside the light cone are canceled by antisymmetry in $t$ due to the choice of a forward propagator (see also the discussion in the general case). Thus, being derivatives of $G_\textrm{R}^{(2)}$, all the matrix elements of the propagator $D(t,\bm{x})$ are characterized by a bulk contribution proportional to $(t^2-r^2)^{-3/2}\theta(t-r)$ and a surface contribution proportional to $(t^2-r^2)^{-1/2}\delta(t-r)$, which compensates the singularity of the first term.
Therefore Huygens' principle is \emph{not} valid, due to the bulk contribution arising from the Heaviside function.

\subsection{3+1 dimensions}
\label{sec:31dim}
In three spatial dimensions, one can use the Weyl representation of the Dirac matrices
\begin{equation}
\gamma^0 = \left( \begin{matrix} 0 & \mathbf{1}_2 \\ \mathbf{1}_2 & 0 \end{matrix} \right) , \quad \gamma^i = \left( \begin{matrix} 0 & \sigma^i \\ -\sigma^i & 0 \end{matrix} \right)
\end{equation}
to decouple the massless Dirac equation into a pair of Weyl equations for two-component spinors. The forward propagators of the upper and lower components are related to each other by Hermitian conjugation. In particular, the upper components are evolved in momentum space by the $2\times 2$ matrix
\begin{equation}
\e^{\i\bm{\sigma}\cdot\bm{p} t} = \left( \begin{matrix} \cos(pt) + \i p^3 \frac{\sin(pt)}{p} & (\i p^1 + p^2) \frac{\sin(pt)}{p} \\ (\i p^1 - p^2) \frac{\sin(pt)}{p} & \cos(pt) - \i p^3 \frac{\sin(pt)}{p}\end{matrix}  \right),
\end{equation} 
that, in position representation, reads
\begin{equation}
D_{\textrm{R}}^{(3,\textrm{up})}(t,\bm{x}) = \left( \begin{matrix} \partial_0 + \partial_3 & \partial_1 -i \partial_2  \\ \partial_1 -i \partial_2 & \partial_0 - \partial_3 \end{matrix}  \right) \i G_\textrm{R}^{(3)}(t,\bm{x}).
\end{equation}
In the $n=3$ case, the integral
\begin{eqnarray}
G_\textrm{R}^{(3)} (t,\bm{x}) & = & -\i \int \frac{\d^3p}{(2\pi)^3} \frac{\sin(pt)}{p} \e^{\i\bm{p}\cdot\bm{x}} = - \frac{2\i}{(2\pi)^2 r} \int_0^{\infty} \d p \sin(pr) \sin(pt)
\end{eqnarray}
determines all the matrix element of the massless Dirac propagator. By observing the symmetry of the integrand in $p=|\bm{p}|$ (again, a peculiar feature of the forward propagator) and excluding the possibilities $t<0$ and $r<0$, one gets
\begin{equation}
G_\textrm{R}^{(3)} (t,\bm{x}) = \frac{-\i}{4\pi r} \delta (t-r).
\end{equation}
In this case, the presence of both $e^{\pm \i p t}$ terms in the integrand cancels the principal value contributions that would arise if only the positive- or negative-energy parts were considered. Integrating by parts, one can observe that the solution at a time $t$ is fixed by the values of the initial spinor components and of their spatial derivatives on the light cone. Huygens' principle is therefore valid.

\section{General considerations on the $n+1$ case}
\label{sec:nplus1}

\begin{figure*}[t]
\centering
\includegraphics[width=0.50\textwidth]{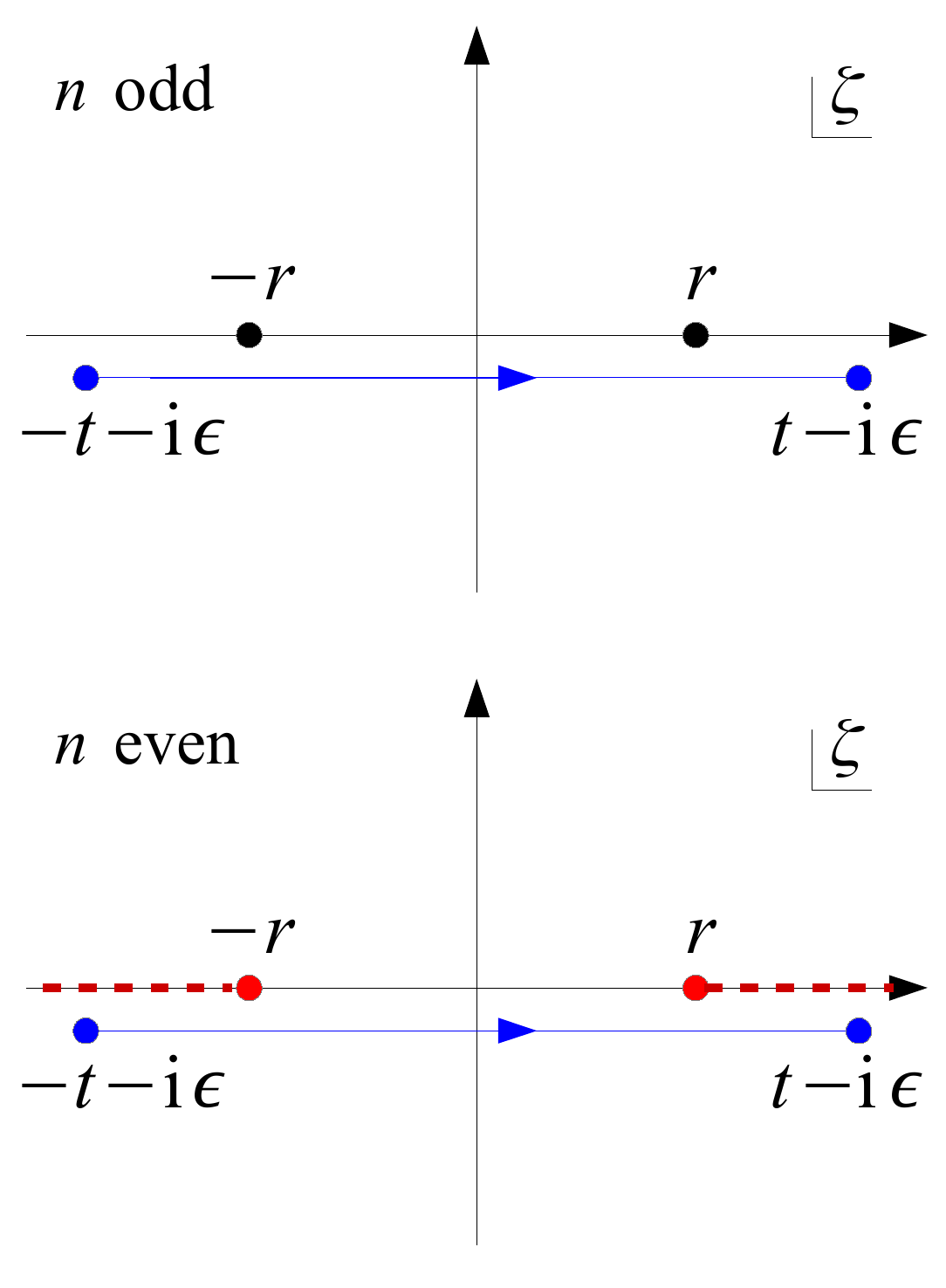}
\caption{Singularities of the propagator on the real axis, for even (top) and odd $n$ (bottom), see Eq.~\eqref{Gzeta}. The arrow in the integration segment (in blue) shows the direction of integration. }
\label{fig:sing}
\end{figure*} 

For general space dimensions $n \geq 2$, the massless KG propagator \label{propKGexpl} can be expressed as an integral over a finite segment in the auxiliary variable $\zeta$ \cite{evans}:
\begin{equation}
\label{Gzeta}
G_\textrm{R}^{(n)} (t,\bm{x}) = 
\theta(t) \frac{\Gamma\left(\frac{n+1}{2}\right)}{\pi^{\frac{n+1}{2}}} \lim_{\varepsilon\to 0^+} \int_{-t-\i\varepsilon}^{t-\i\varepsilon} \d\zeta \frac{\zeta}{(r^2-\zeta^2)^{\frac{n+1}{2}}} ,
\end{equation}
where $\Gamma$ is the Gamma function and $r=|\bm{x}|$. 
(This formula can also be written for $n=1$, but care must be taken in handling the singularities at $r=0$.)
This expression, plugged into Eq.\ (\ref{dpropform}), yields the Dirac forward propagator.
For arbitrary $n$, if $r>t$, the integrand has no singularity on the real axis, and the propagator vanishes by symmetry as $\varepsilon\to 0$. The different nature of the singularities on the segment $[-t,t]$ determines the ``locality" of the propagator: for odd $n$, two poles are present at $\zeta=\pm r$, while for even $n$, $\zeta=\pm r$ are the branching points of two cuts on the real axis, that extend to $\pm\infty$. The different behavior is represented in Fig.~\ref{fig:sing}. 
The action of $\slashed{\partial}$ in Eq.\ (\ref{dpropform}) does not modify this fundamental property, and clarifies why Huygens' is valid in the massless Dirac equation for odd $n (\geq3)$, but it is not valid for even $n$. 

The case $n=1$ stands out as an important difference. In this case, it is most convenient to start from Eq.\ (\ref{propKGR}), to obtain
\begin{eqnarray}
G_\textrm{R}^{(1)} (t,x) &=& 
\theta(t) \int \frac{\d p}{2 \pi}
\int_{-t}^{t} \frac{\d\zeta}{2\i} \e^{-\i p (\zeta-x)} = -\frac{\i}{2} \theta(t) [\theta(t-x) - \theta(-t-x)].
\label{Gzeta1}
\end{eqnarray}
As one can see, the massless KG propagator has bulk contributions. \emph{However}, the action of $\slashed{\partial}$ (that involves $\partial_0$ and $\partial_1$) transforms them in boundary terms, as in Eq.\ (\ref{D1}), restoring the validity of Huygens' principle. This remarkable circumstance does not occur for even $n$. In this case the theta function is multiplied by $(t,\bm{x})$-dependent factors, as in Eq.\ (\ref{GR2}), that make the bulk contribution survive the action of the derivatives, making Huygens' principle invalid.

\section{Conclusions}
\label{sec:nplus1}
In this Article we discussed the features of propagation for the massless Dirac-Weyl equation. We found that Huygens' principle is valid for odd spatial dimensions, while it is not valid for even spatial dimensions. 
Interestingly, the principle remains valid even for the case $n = 1$ (at variance with the Klein-Gordon case). This is due to the fact that the relations among the spinorial components of the wave function impose extra conditions on the initial-value problem, that enforce Huygens' principle to be valid also in this case.

We worked with retarded boundary conditions for the Green function, propagating solutions \emph{forward} in time. This option is particularly suited for the investigation of electron motion in graphene, close to Dirac's points, where the description via a two-component spinor wave function in two space dimensions is very effective \cite{graphene_rev,graphene_nat}. 
There are other interesting applications that one can consider, such as the quantum simulations of QED and in general lattice gauge theories and low-dimensional quantum systems \cite{qsim1,qsim2,qsim3,qsim4}. The differences between, say, $n=1$ and $2$, might display interesting signatures of bulk \emph{vs} boundary effects in the propagation of physical observables.

There are a number of problems that one can investigate in the future. One of these is the extension of these ideas to the massive Dirac equation. Another interesting problem would be to unveil the effects of dimensionality when an external field is introduced via the minimal coupling prescription. In general, this bears consequences on the way the coupling to external classical fields is handled. Finally, an important open question is the relationship among Huygens' principle, dimensionality and Feynman's prescription for the propagator.

\section*{Acknowledgments}
We thank Alessandro Zampini for interesting discussions.
FVP and SP are partly supported by INFN through the projects ``PICS" and ``QUANTUM". 
JMPP is partly supported by the Spanish MINECO grant MTM2014-54692-P and QUITEMAD+, S2013/ICE-2801.


\begin{thebibliography}{}

\bibitem{BornWolf}
M. Born and E. Wolf, \textit{Principles of Optics} (Cambridge University Press, Cambridge, 1999).

\bibitem{Hadamard}
J. Hadamard, \textit{Lectures on Cauchy's Problem in Linear Partial Differential Equations} (Dover Publications, New York, 1923).

\bibitem{zac}
E. C. Zachmanoglou and D. W. Thoe, \textit{Introduction to Partial Differential Equations with Applications} (Dover Publications, New York, 1986).

\bibitem{evans}
L. C. Evans, \textit{Partial Differential Equations}, Graduate Studies in Mathematics, Vol. 19 (American
Mathematical Society, Providence, Rhode Island, 2001).

\bibitem{Gunther}
P. G\"unther, \textit{Huygens' principle and hyperbolic equations} (Academic Press, San Diego, 1988). 

\bibitem{chalub}
F. A. C. C. Chalub and J. P. Zubelli, J. Nonlinear Math. Phys., \textbf{8}, 62 (2001).

\bibitem{chalub2}
F. A. C. C. Chalub and J. P. Zubelli, 
Cont. Math. AMS \textbf{362}, 89 (2004).

\bibitem{Peskin}
M. E. Peskin and D. V. Schroeder, \textit{An introduction to Quantum Field Theory} (Perseus Books, Reading, 1995).

\bibitem{Feynman}
R. P. Feynman, \textit{Quantum Electrodynamics} (W. A. Benjamin, New York, 1961).

\bibitem{graphene_rev}
A. H. Castro Neto, F. Guinea, N. M. R. Peres, K. S. Novoselov, and A. K. Geim, Rev. Mod. Phys \textbf{81}, 109 (2009).

\bibitem{graphene_nat}
M. I. Katsnelson, K. S. Novoselov, and A. K. Geim, Nat. Phys. \textbf{2}, 620 (2006).

\bibitem{qsim1}
E. Zohar, J. I. Cirac, and B. Reznik, Phys. Rev. Lett. \textbf{109}, 125302 (2012).

\bibitem{qsim2}
D. Banerjee, M. Dalmonte, M. M\"uller, E. Rico, P. Stebler, U.-J. Wiese, and P. Zoller, Phys Rev. Lett \textbf{109}, 175302 (2012).

\bibitem{qsim3}
S. Notarnicola, E. Ercolessi, P. Facchi, G. Marmo, S. Pascazio, and F. V. Pepe, 
J. Phys. A: Math. Theor. \textbf{48}, 30FT01 (2015). 

\bibitem{qsim4}
T. Pichler, M. Dalmonte, E. Rico, P. Zoller, and S. Montangero, Phys. Rev. X \textbf{6}, 011023 (2016).

\end{thebibliography}
\end{document}